\documentclass[final,3p,times]{elsarticle}
\usepackage{amssymb}
\usepackage{amsmath}
\journal{Journal of Subatomic Particles and Cosmology}

\begin{document}
\begin{frontmatter}

\title{Recent STAR Measurements from the RHIC Beam Energy Scan II}

\author[lbnl]{Ashish Pandav\corref{cor1} (for the STAR Collaboration)} 

\cortext[cor1]{e-mail: ashishp.pandav@gmail.com}

\address[lbnl]{Lawrence Berkeley National Laboratory, Berkeley, CA 94720, USA}

\begin{abstract}
We present selected recent results from the second phase of the Beam Energy
Scan program (BES-II) at RHIC, based on high-statistics data collected by the
STAR experiment in collider and fixed-target modes, covering
$\sqrt{s_{NN}}=3.0$--$27$~GeV. New results from isobar (Ru+Ru, Zr+Zr)
collisions at $\sqrt{s_{NN}} = 200$~GeV are also presented. The presented results span bulk properties and strangeness production, femtoscopy, collectivity and
hyperon polarization, and fluctuation observables sensitive to the QCD phase
structure. Future prospects with newly recorded fixed-target datasets
are also briefly outlined.
\end{abstract}


\end{frontmatter}

\section{Introduction}

Mapping the phase structure of quantum chromodynamics (QCD) in the plane of
temperature $T$ and baryon chemical potential $\mu_B$ is a central goal of
relativistic heavy-ion physics~\cite{Bzdak:2019pkr,Pandav:2022}. Lattice QCD
establishes a smooth crossover between hadronic matter and the quark--gluon
plasma (QGP) at $\mu_B\simeq0$, while QCD-based model calculations predict a
first-order phase boundary at large $\mu_B$ terminating in a critical point
(CP). Since the chemical freeze-out parameters $(T,\mu_B)$ extracted from
hadron yields vary systematically with collision
energy~\cite{Pandav:2022,BraunMunzinger:2007}, a beam energy scan effectively
probes a wide range of the QCD phase diagram along the freeze-out curve.

The second phase of the Beam Energy Scan (BES-II) at the
Relativistic Heavy Ion Collider (RHIC) delivered high-statistics Au+Au data at
$\sqrt{s_{NN}}=7.7$--$27$~GeV in collider mode and down to
$\sqrt{s_{NN}}=3.0$~GeV in the STAR fixed-target (FXT) program, giving STAR
coverage of $3\le\sqrt{s_{NN}}\,\mathrm{(GeV)}\le200$, i.e.,
$25\lesssim\mu_B\,\mathrm{(MeV)}\lesssim750$. Event samples are roughly an order of magnitude larger than in BES-I, while the inner Time Projection Chamber
(iTPC), endcap Time-Of-Flight (eTOF), and Event Plane Detector (EPD) upgrades
provide extended acceptance, lower-$p_T$ tracking, improved particle
identification, and improved event-plane resolution. RHIC completed
its final run in 2026, concluding a quarter century of STAR operations. Selected recent results from BES-II are discussed in this contribution.
STAR highlights on small systems and exotic physics are presented in Ref.~\cite{Jia:SQM2026}.
\section{Bulk properties and strangeness}
\label{sec:bulk}

\paragraph{Hadron spectra and freeze-out systematics}
Identified hadron ($\pi$, $K$, $p$) transverse-mass spectra have been measured
differentially in rapidity and centrality in Au+Au FXT collisions at
$\sqrt{s_{NN}}=3.2$--$4.5$~GeV~\cite{Labonte:SQM2026}. Blast-wave fits are applied to extract the kinetic freeze-out temperature $T_{\rm kin}$ and the mean
transverse expansion velocity $\langle\beta_T\rangle$; both are found to
increase with collision energy, smoothly extending towards the trend
established by BES-I collider data from $\sqrt{s_{NN}}=7.7$ to 200~GeV. In
isobar collisions (Ru+Ru, Zr+Zr) at $\sqrt{s_{NN}}=200$~GeV, chemical
($T_{\rm chem}$, $\mu_B$) and kinetic freeze-out parameters are extracted from
$\pi$, $K$, $p$, $\Lambda$, and $\Xi$ yields~\cite{Tsang:SQM2026}. The
parameters follow the same systematic trends as observed in Au+Au and Cu+Cu collisions at the same energy when plotted as a function of the mean number of participating nucleons, indicating that they are primarily driven by
the size of the produced system rather than the colliding species.

\begin{figure}[!htb]
\centering
\includegraphics[scale=0.22]{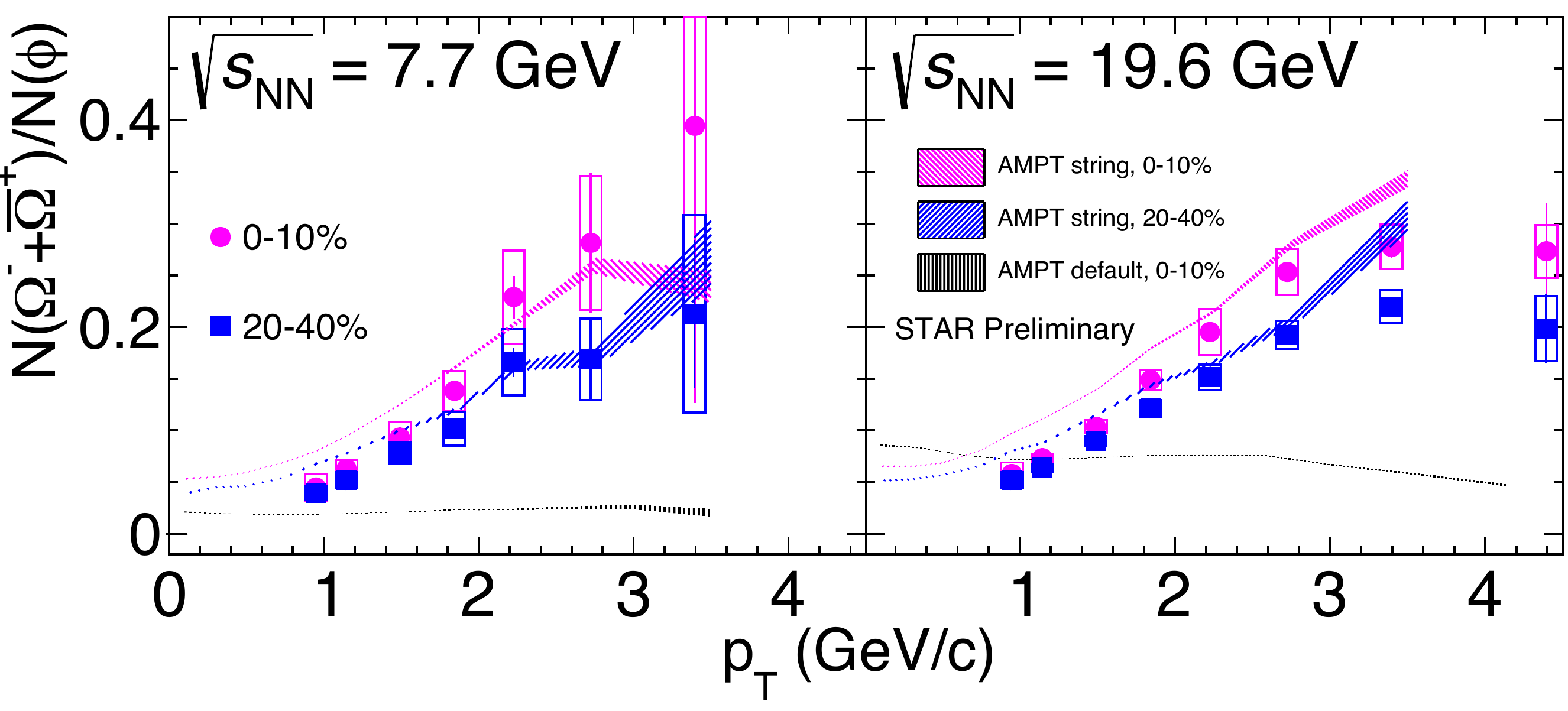}
\caption{Yield ratio $N(\Omega^-+\bar{\Omega}^+)/N(\phi)$ as a function of
$p_T$ measured in Au+Au collisions at $\sqrt{s_{NN}}=7.7$ and 19.6~GeV for
0--10\% and 20--40\% centralities. The bands represent calculations from the
AMPT transport model.}
\label{label_fig1}
\end{figure}

\paragraph{(Multi-)strange hadron production}
Multi-strange hadron yields are sensitive probes of QGP formation
and of the hadronization mechanism.
Figure~\ref{label_fig1} presents the ratio
$N(\Omega^-+\bar{\Omega}^+)/N(\phi)$ measured in Au+Au collisions at
$\sqrt{s_{NN}}=7.7$ and 19.6~GeV, which increases strongly with $p_T$. The default AMPT model, based on string fragmentation, fails to
describe the data, whereas the string-melting version, which implements
QGP-like quark coalescence, reproduces the observed experimental trend~\cite{Yuan:SQM2026}.

\paragraph{Baryon-strangeness correlations}
The correlation between baryon and strangeness numbers, $C_{BS}$, has been
proposed as a sensitive diagnostic of the effective degrees of freedom of QCD
matter~\cite{Asakawa:2000,Koch:2005}. It has been measured using $p$, $K$,
and $\Lambda$ as proxies for baryon number and strangeness in isobar collisions at $\sqrt{s_{NN}}=200$~GeV. The $C_{BS}$ values are found to lie above the hadronic model UrQMD and are
consistent with lattice QCD expectations within uncertainties across the
measured centrality range~\cite{Li:SQM2026}.

\section{Femtoscopy and hypernuclei}
\label{sec:femto}
\begin{figure}[!htb]
\centering
\begin{minipage}[t]{0.48\textwidth}
  \centering
    \includegraphics[scale=0.2]{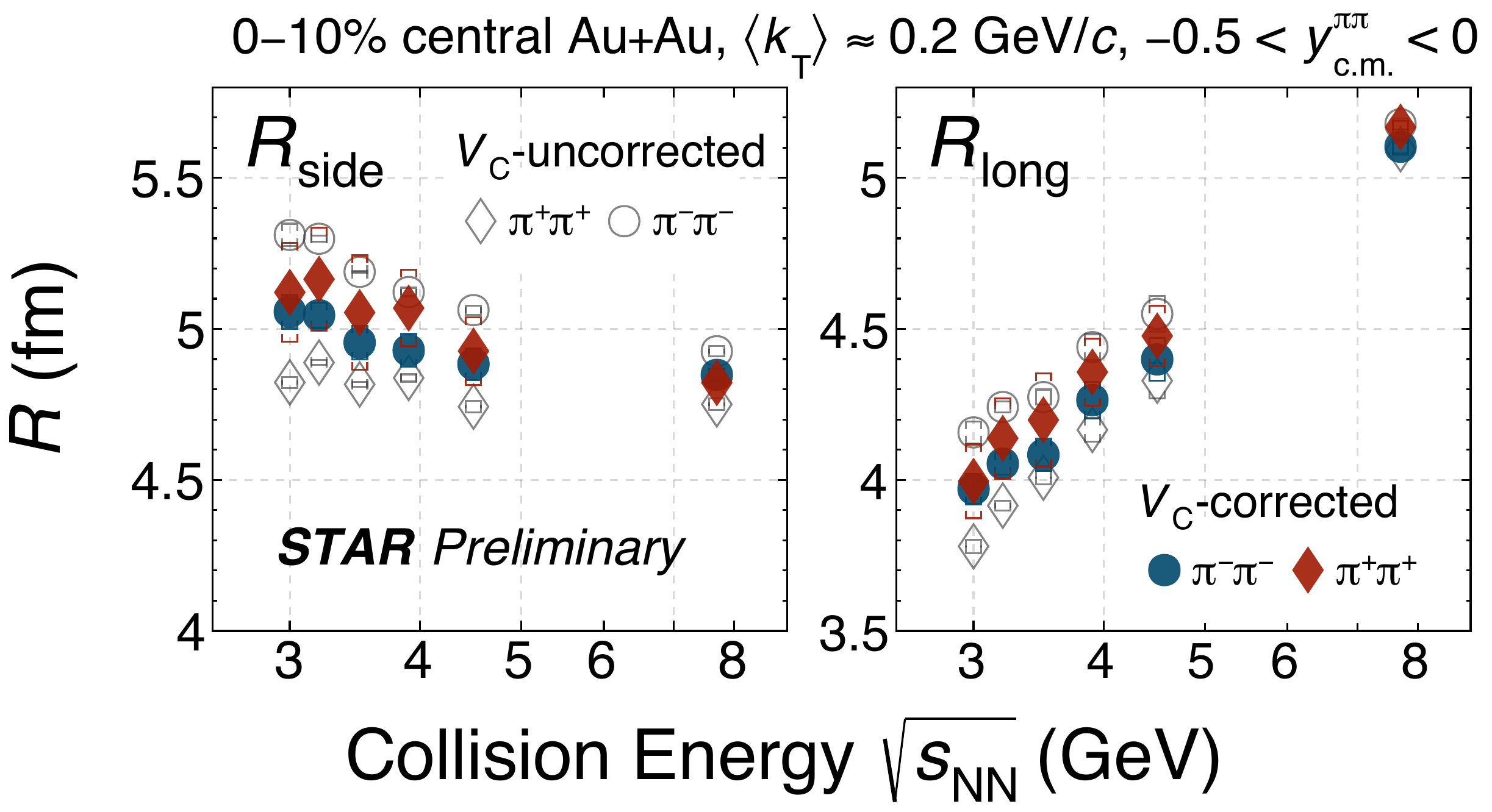}
  \caption{Source radii $R_{\rm side}$ and $R_{\rm long}$ extracted from $\pi^+\pi^+$ and $\pi^-\pi^-$ pairs in 0--10\% central
  Au+Au collisions, before and after the third-body Coulomb correction.}
  \label{label_fig2}
\end{minipage}\hfill
\begin{minipage}[t]{0.48\textwidth}
  \centering
  \includegraphics[scale=0.24]{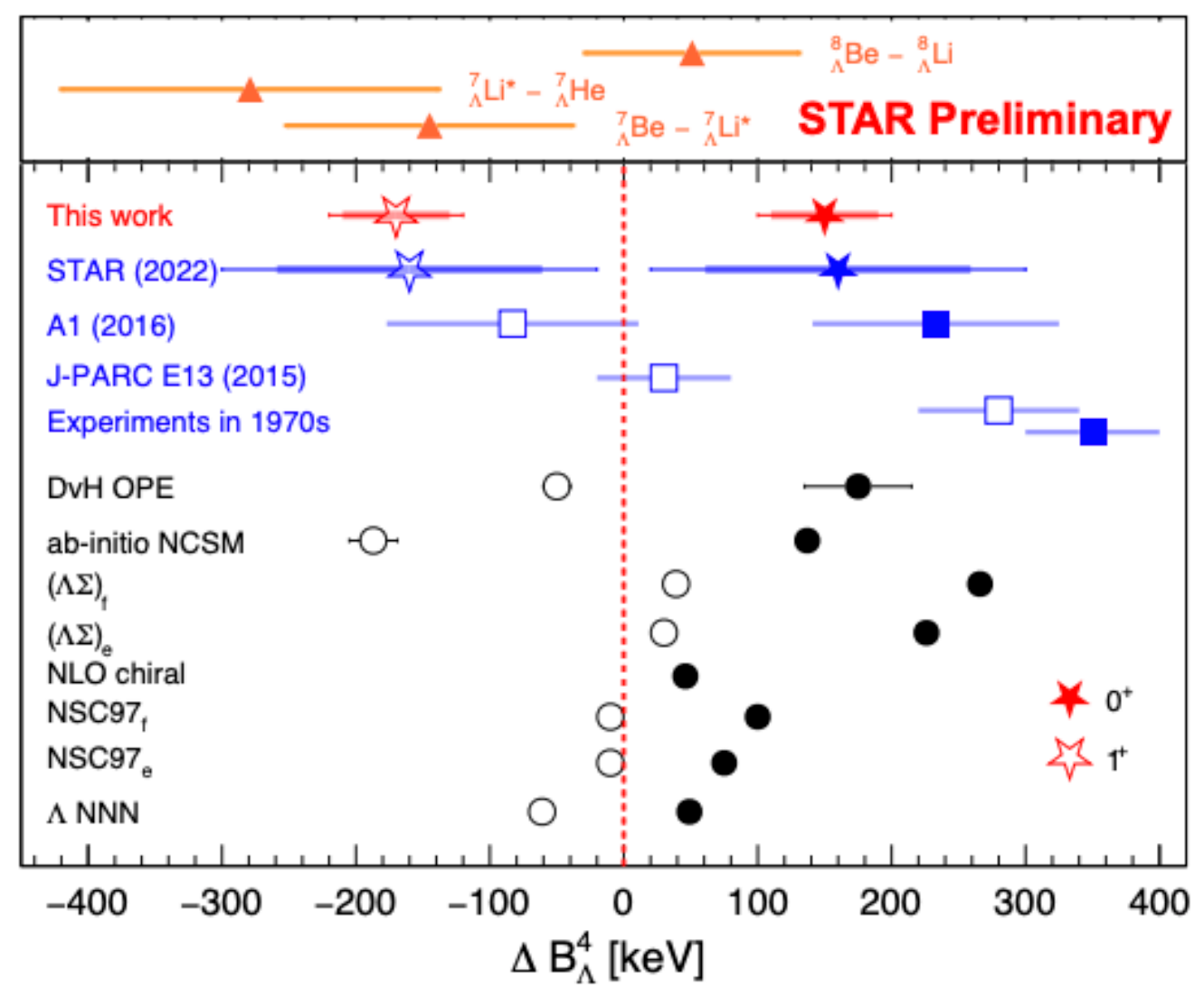}
  \caption{Difference in $\Lambda$ binding energy between $^4_\Lambda$H and
  $^4_\Lambda$He hypernuclei measured in Au+Au collisions at
  $\sqrt{s_{NN}}=3$~GeV for both ground and excited states. Theoretical
  calculations are presented as black markers.}
  \label{label_fig3}
\end{minipage}
\end{figure}

Femtoscopic correlations, $C(k^*)=\int dr^*\,S(r^*)\,|\Psi(r^*,k^*)|^2$,
simultaneously probe the space--time geometry of the emitting source, encoded
in the source function $S(r^*)$, and the final-state interaction, encoded in
the pair wave function
$\Psi(r^*,k^*)$~\cite{Lednicky:1982,Lisa:2005}.

\paragraph{Identical-pion correlations and the third-body Coulomb effect}
In Au+Au collisions at FXT energies, the source radii $R_{\rm side}$ and
$R_{\rm long}$ extracted from $\pi^+\pi^+$ pairs are systematically smaller
than those from $\pi^-\pi^-$ pairs, as seen in Fig.~\ref{label_fig2} for 0--10\% central collisions~\cite{Qi:SQM2026}. Applying a correction
based on the Coulomb potential energy ($V_c$) extracted from charged-pion
$p_T$ spectra brings the $\pi^+\pi^+$ and $\pi^-\pi^-$ radii into agreement
across $\sqrt{s_{NN}}=3.0$--7.7~GeV, demonstrating that the third-body Coulomb
interaction from the residual Coulomb field of the positively charged fireball
is the dominant cause of the observed charge splitting.

\paragraph{Hyperon--nucleon interactions from $p$-$\Xi^-$ and
$p$-$p$-$\Lambda$ correlations}
Correlations of protons with $\Xi^-$ hyperons have been measured in Au+Au
collisions at $\sqrt{s_{NN}}=7.7$, 14.6, and 19.6~GeV, exhibiting a clear
enhancement above the Coulomb-only expectation~\cite{An:SQM2026}. Scattering
parameters extracted with the Lednick\'y--Lyuboshitz formalism reveal a
hierarchy of the scattering length with strangeness content,
$f_0(|s|=0)>f_0(|s|=1)>f_0(|s|=2)>0$, when compared with $p$-$p$ and
$p$-$\Lambda$ results at 3~GeV. Going beyond two-body systems, STAR has also
performed its first measurement of the genuine three-body $p$-$p$-$\Lambda$
correlation function in Au+Au collisions at $\sqrt{s_{NN}}=3$~GeV, with good
kinematic coverage down to relative momenta below 100~MeV, where sensitivity
to final-state interactions is largest~\cite{Gu:SQM2026}. The comparisons with theoretical calculations emphasize the need for further development of
the modeling of the three-body force~\cite{Garrido:2024}.

\paragraph{Charge symmetry breaking in $A=4$ hypernuclei}
The mirror hypernuclei $^4_\Lambda$H and $^4_\Lambda$He provide a sensitive
probe of the isospin dependence of the $\Lambda N$ interaction. Using Au+Au
collision data from the BES-II FXT program at $\sqrt{s_{NN}}=3$~GeV, STAR has
measured the difference in $\Lambda$ binding energies, $\Delta B^4_\Lambda$,
~\cite{STAR:2026csb,Shao:SQM2026}. The result,
presented in Fig.~\ref{label_fig3}, constitutes the most precise
determination of charge symmetry breaking (CSB) in the $A=4$ hypernuclear
system to date. The CSB strength is found to be of similar magnitude in the
ground ($0^+$) and excited ($1^+$) states within uncertainties but with
opposite sign, providing constaints for ab initio and chiral effective
field theory descriptions of the YN force.

\section{Collectivity and hyperon polarization}
\label{sec:flow}
\begin{figure}[!htb]
\centering
\begin{minipage}[t]{0.48\textwidth}
  \centering
      \includegraphics[scale=0.2]{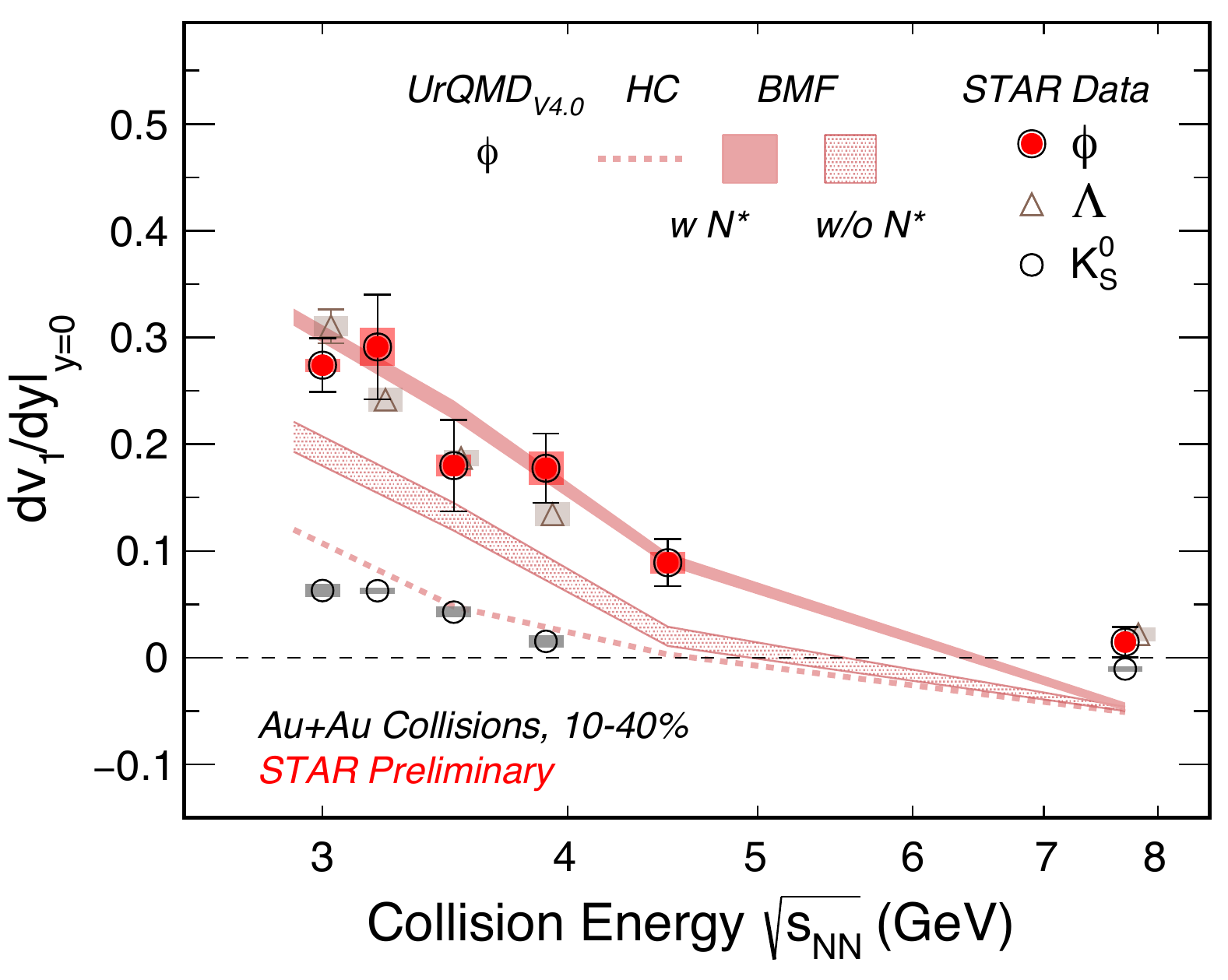}
  \caption{Directed-flow slope $dv_1/dy|_{y=0}$ of $\phi$, $\Lambda$, and
  $K^0_S$ in 10--40\% Au+Au collisions as a function of $\sqrt{s_{NN}}$.
  Results are compared to UrQMD calculations in hadronic-cascade (HC) and
  baryonic-mean-field (BMF) modes, with and without baryon-resonance ($N^*$)
  excitations.}
  \label{label_fig4}
\end{minipage}\hfill
\begin{minipage}[t]{0.48\textwidth}
  \centering
      \includegraphics[scale=0.20]{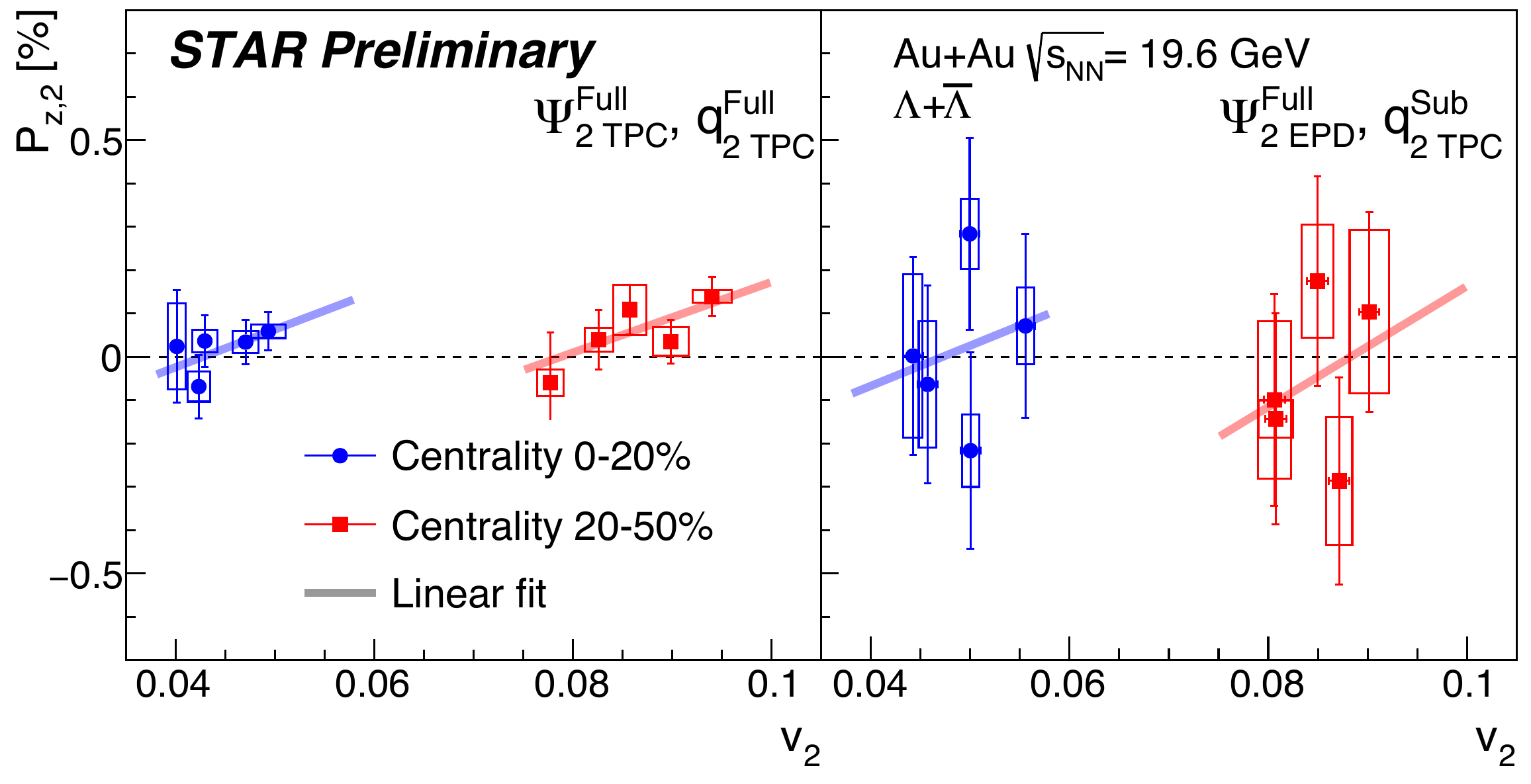}
  \caption{Local polarization harmonic $P_{z,2}$ of $\Lambda+\bar{\Lambda}$
  vs elliptic flow $v_2$ in Au+Au collisions at $\sqrt{s_{NN}}=19.6$~GeV for
  two centrality classes, using the event plane $\Psi_2$ and flow vector
  $q_2$ from the full TPC (a) and the EPD event plane with a TPC sub-event
  $q_2$ (b); lines are linear fits.}
  \label{label_fig5}
\end{minipage}
\end{figure}
\paragraph{Directed flow of the $\phi$ meson at high baryon density}
The $\phi$ meson, with its small hadronic cross section and comparatively long
lifetime, is a clean probe of the early stage of the collision. As seen in
Fig.~\ref{label_fig4}, STAR has measured a large directed-flow slope
$dv_1/dy|_{y=0}$ for the $\phi$ in 10--40\% Au+Au collisions at
$\sqrt{s_{NN}}=3.0$--7.7~GeV, comparable to that of the $\Lambda$ and the proton (not shown in Fig. 4), which is much larger than that of
$K^0_S$~\cite{Zheng:SQM2026}. UrQMD calculations including a baryonic mean
field reproduce the measurements only when baryon-resonance excitation
channels ($N+N\to N^*+N\to N+\phi+N$) are included, establishing the $\phi$
meson as a sensitive probe of baryon-resonance dynamics in the
high-baryon-density region.

\paragraph{Global and local polarization of hyperons}
Non-central heavy-ion collisions also provide access to the rotational
properties of the medium through the conversion of the system's huge orbital
angular momentum into vorticity, observable via hyperon spin polarization
measurements. With BES-II statistics, STAR reports significant global
polarization of $\Xi^-+\bar{\Xi}^+$, with
$P_H(\Xi^-+\bar{\Xi}^+)\simeq P_H(\Lambda+\bar{\Lambda})$ within uncertainties
and a hint of larger $\Omega^-+\bar{\Omega}^+$ polarization, decreasing with
collision energy and in line with AMPT-based model
expectations~\cite{Fu:SQM2026,Li:2022}. In addition, first measurements of the
correlation between the $\Lambda+\bar{\Lambda}$ local (longitudinal)
polarization harmonic $P_{z,2}$ and the elliptic flow $v_2$ have been
performed at $\sqrt{s_{NN}}=19.6$~GeV using event-shape
engineering~\cite{Kondo:SQM2026}, shown in Fig.~\ref{label_fig5}.
A positive dependence of $P_{z,2}$ on the flow vector $q_2$ and $v_2$ is
suggested, although the effect is seen at a level of ${\lesssim}2\sigma$ with the current uncertainties.
This constitutes an initial step toward experimentally testing the
hydrodynamic origin of local vorticity.

\section{Fluctuations and the search for the QCD critical point}
\label{sec:cp}

\begin{figure}[!htb]
\centering
\begin{minipage}[t]{0.48\textwidth}
  \centering
        \includegraphics[scale=0.215]{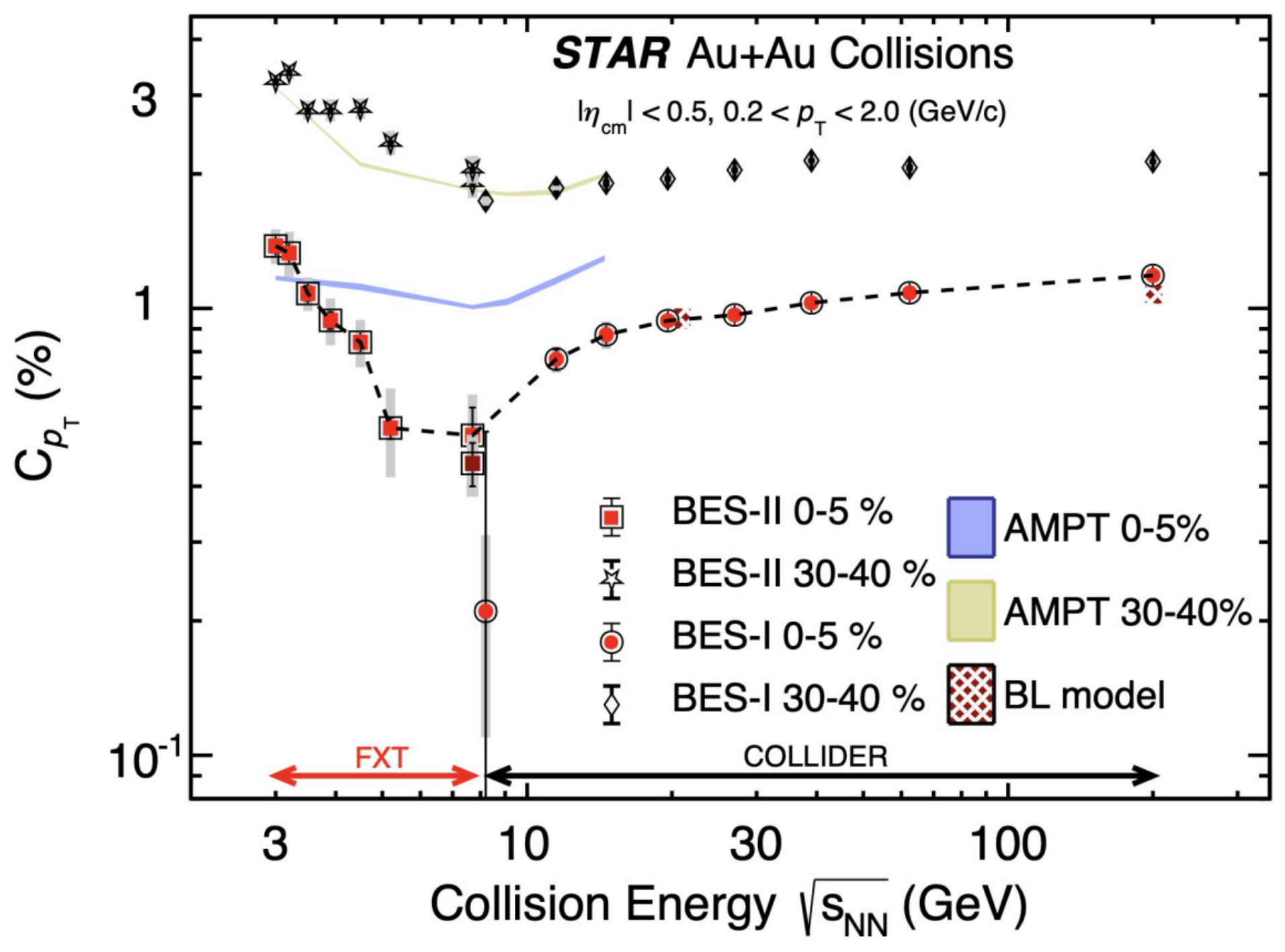}
  \caption{Mean-$p_T$ correlator $C_{p_T}$ in 0--5\% and 30--40\% Au+Au
  collisions, compared to the transport model AMPT and Boltzmann--Langevin
  (BL) model calculations~\cite{Gavin:2017}. Bars and shaded bands denote
  statistical and systematic uncertainties, respectively.}
  \label{label_fig6}
\end{minipage}\hfill
\begin{minipage}[t]{0.48\textwidth}
  \centering
  \includegraphics[scale=0.22]{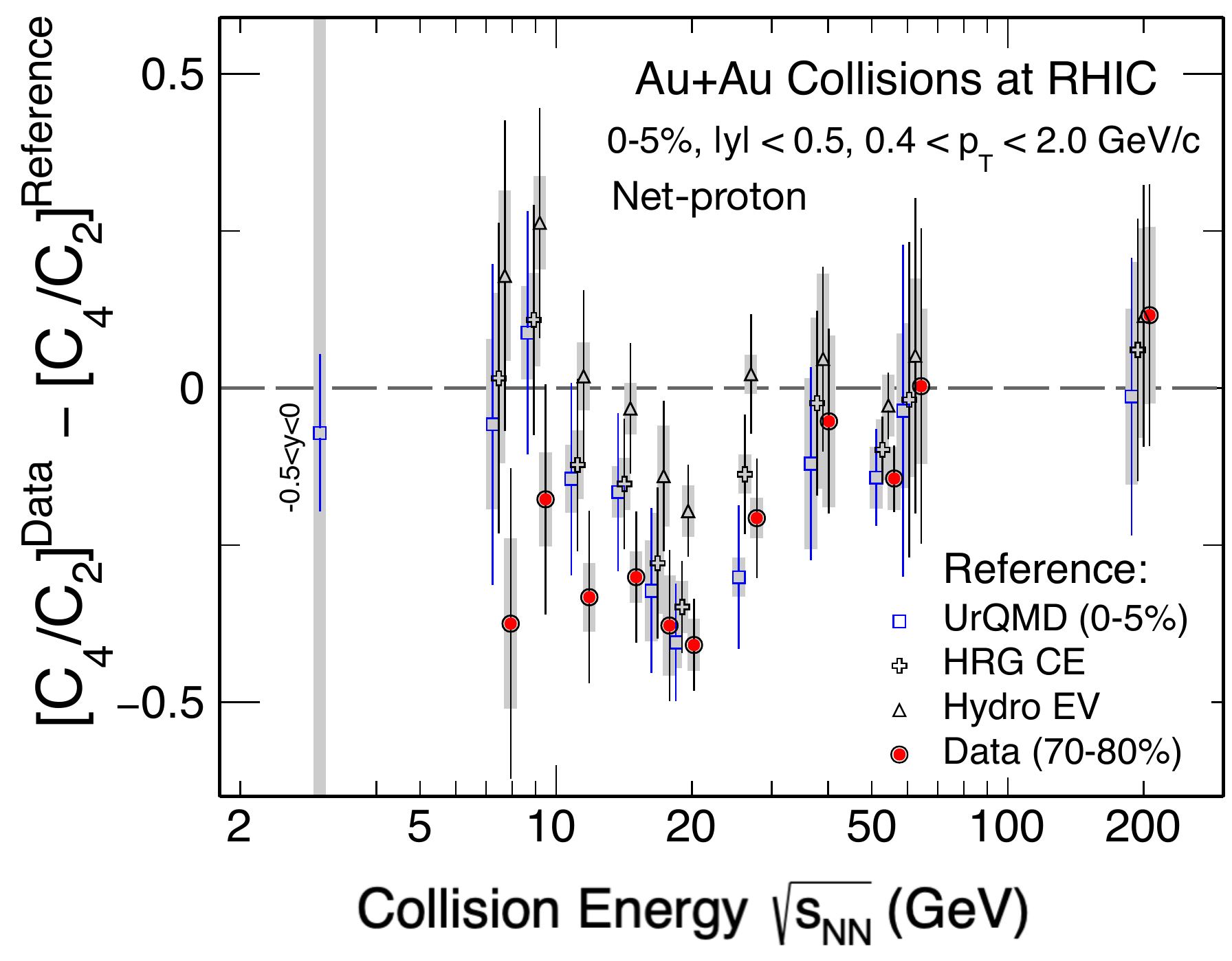}
  \caption{Deviation of net-proton $C_4/C_2$ in 0--5\% central Au+Au
  collisions from non-critical references such as UrQMD, HRG with canonical
  ensemble (CE), hydrodynamical model calculations with excluded volume (EV)
  effects, and peripheral 70--80\% data, as a function of
  $\sqrt{s_{NN}}$~\cite{STAR:2025}.}
  \label{label_fig7}
\end{minipage}
\end{figure}

In the vicinity of a critical point, the correlation length of the system
grows, and event-by-event fluctuations of charged-particle multiplicity and
momentum are predicted to develop non-monotonic dependencies on collision
energy~\cite{Stephanov:2011}. The precision of BES-II, together with the FXT
extension to $\mu_B\approx750$~MeV, makes this the flagship physics of the
program.

\paragraph{Mean-$p_T$ fluctuations}
The intensive correlator
$C_{p_T}=\sqrt{\langle\Delta p_{T,i}\Delta p_{T,j}\rangle}/\langle\langle
p_T\rangle\rangle$, robust against volume and efficiency effects, has recently
been measured across the STAR FXT energy
range~\cite{STAR:2026cpt,Manikandhan:SQM2026}, extending the earlier measurements in the
collider region. In 0--5\% central Au+Au collisions, $C_{p_T}$ exhibits a
non-monotonic collision energy dependence with a minimum around
$\sqrt{s_{NN}}=4.5$--7.7~GeV, as shown in Fig.~\ref{label_fig6}. Neither the AMPT model, which contains no phase transition, nor the 30--40\%
mid-central data exhibit such pronounced structure. Since mean-$p_T$
fluctuations are sensitive to the equation of state, and hence to a possible
first-order phase transition or critical point, this observation is
particularly intriguing.

\paragraph{Net-proton higher-moment cumulants}
Higher-order cumulants of the net-proton multiplicity distribution are the
most direct experimental proxy for baryon-number susceptibilities. Near a
critical point, the ratio $C_4/C_2$ is predicted to dip below and then rise
above its non-critical baseline as one moves from high to low collision
energies~\cite{Stephanov:2011}. In 0--5\% central Au+Au collisions, BES-II
data show a minimum in net-proton $C_4/C_2$ with respect to non-critical
reference calculations (hadronic model UrQMD, hydrodynamical model with excluded volume
effects, HRG with canonical ensemble) and to peripheral (70--80\%) data, at a
significance of 2--5$\sigma$ depending on the baseline, around
$\sqrt{s_{NN}}\approx20$~GeV, as shown in
Fig.~\ref{label_fig7}~\cite{STAR:2025}. For $\sqrt{s_{NN}}\ge27$~GeV, the
data are consistent with the non-critical baselines. In addition, the analyzed
FXT data extend the measurement down to
$\sqrt{s_{NN}}=3.0$--3.9~GeV~\cite{STAR:2022cum3gev,Sweger:QM2025}, where consistency with the
hadronic baseline UrQMD is observed.

\section{Summary and outlook}
\label{sec:summary}

The STAR BES-II program has been delivering precision measurements across the key
frontiers of QCD at high baryon density. Freeze-out systematics now extend
smoothly from top RHIC energy down to $\sqrt{s_{NN}}=3.2$~GeV, and strangeness
observables ($\Omega/\phi$, $C_{BS}$) consistently indicate partonic degrees
of freedom for $\sqrt{s_{NN}}\gtrsim7.7$~GeV. Femtoscopic studies have
advanced new information on the hyperon--nucleon interactions, and
measurements of charge symmetry breaking in $A=4$ hypernuclei have been
improved significantly. Hyperon polarization studies have progressed to
detailed measurements of multi-strange hyperon polarization across collision
energies and of the correlation between local polarization and elliptic flow.

On the flagship goal of the program, the search for the QCD critical point,
mean-$p_T$ fluctuations show a non-monotonic energy dependence with a minimum
around $\sqrt{s_{NN}}=4.5$--$7.7$~GeV. In addition, the net-proton $C_4/C_2$ exhibits a minimum near
$\sqrt{s_{NN}}\approx20$~GeV with respect to baselines without a QCD critical
point, at a significance of 2--5$\sigma$. Such a minimum is one of the
characteristic features of a proposed critical-point signal. Dynamical model
calculations including a critical point are needed for a quantitative
interpretation of the observed pattern, and $C_4/C_2$ measurements in the
region $3.9<\sqrt{s_{NN}}<7.7$~GeV will be key to the continued search. This
is especially so given that several recent theoretical estimates place the critical point around
$\sqrt{s_{NN}}\sim4$--$5$~GeV
($\mu_B\approx550$--$650$~MeV)~\cite{Clarke:2024,Basar:2024,Hippert:2024,Fu:2020,Gunkel:2021,Gao:2021,Sorensen:2024,Huang:QM2025,Shah:2026}.

In its final runs, STAR recorded large FXT datasets at
$\sqrt{s_{NN}}=4.2$~GeV (${\sim}290$ million events), 4.5~GeV (${\sim}1$ billion), and
5.2~GeV (${\sim}370$ million), enabling precision fluctuation measurements in
exactly this region. Beyond fluctuations, these data will also enable searches
for new physics, such as a possible observation of double-$\Lambda$
hypernuclei via
$^4_{\Lambda\Lambda}\mathrm{H}\to{}^4_\Lambda\mathrm{He}+\pi^-$, whose
production is predicted to peak in this energy
range~\cite{Andronic:2011,Steinheimer:2012}. Together with the wealth of
BES-II data already recorded, these datasets will keep the exploration of QCD
at high baryon density active for years to come.

\section*{Acknowledgments}
The author thanks the SQM2026 organizers. This work is supported by
the U.S. Department of Energy, Office of Science.


\end{document}